\documentclass[conference,letterpaper]{IEEEtran}

\usepackage[utf8]{inputenc} 
\usepackage[T1]{fontenc}
\usepackage{url}
\usepackage{ifthen}
\usepackage{cite}
\usepackage[cmex10]{amsmath} 
\usepackage{xcolor}
\usepackage{tabularx,ragged2e,booktabs}
\usepackage{times}
\usepackage{float}
\usepackage{afterpage}
\usepackage{amstext}
\usepackage{amssymb,bm}
\usepackage{latexsym}
\usepackage{color}
\usepackage{pstricks}
\usepackage{enumerate}
\usepackage{mathtools}
\usepackage{epsfig}
\usepackage{afterpage}
\usepackage{amstext}
\usepackage{amssymb}
\usepackage{amssymb,bm}
\usepackage{latexsym}
\usepackage{color}
\usepackage{graphicx}
\usepackage{amsthm}
\usepackage{graphicx}
\usepackage{pstricks}
\usepackage{booktabs}
\usepackage{enumerate}
\usepackage{subfig}
\usepackage{fixltx2e}
\usepackage{enumitem}
\usepackage{multirow}
\usepackage{verbatim}
\usepackage{mathabx}
\usepackage{hyperref}

\DeclarePairedDelimiter\ceil{\lceil}{\rceil}

\newtheorem{thm}{Theorem}

\newcommand{\smin}{s_{\rm min}} 
\newcommand{\smax}{s_{\rm max}} 
\newcommand{\cardi}{t} 
\newcommand{\mbit}{\kappa} 
\newcommand{\fspli}{f} 
\newcommand{\Mod}[1]{\ (\mathrm{mod}\ #1)}

\begin{document}
\title{Decentralized Pliable Index Coding} 

\author{%
\IEEEauthorblockN{%
Tang Liu and Daniela Tuninetti\\%
University of Illinois at Chicago, Chicago, IL 60607 USA, \\
Email: {\tt tliu44, danielat@uic.edu}\\%
}%
}
\maketitle




\begin{abstract}
This paper introduces the {\it decentralized} Pliable Index CODing (PICOD) problem: a variant of the Index Coding (IC) problem, where a central transmitter serves {\it pliable} users with message side information; here, pliable refers to the fact that a user is satisfied by decoding {\it any} $\cardi$ messages that are not in its side information set.
In the decentralized PICOD, a central transmitter with knowledge of all messages is not present, and instead users share among themselves massages that can only depend on their local side information set. This paper characterizes the capacity of two classes of decentralized \emph{complete--$S$} PICOD$(\cardi)$ problems with $m$ messages (where the set $S\subset[m]$ contains the sizes of the side information sets, and the number of users is $n=\sum_{s\in S}\binom{m}{s}$, with no two users having the same side information set):
(i)  the \emph{consecutive case} $S=[\smin:\smax]$ for some $0 \leq \smin\leq \smax \leq m-\cardi$, and
(ii) the \emph{complement-consecutive case} $S=[0:m-\cardi]\backslash[\smin:\smax]$, for some $0 < \smin\leq \smax < m-\cardi$.
Interestingly, the optimal code-length for the decentralized PICOD in those cases is the same as for the classical (centralized) PICOD counterpart, except when the problem is no longer pliable, that is, it reduces to an IC problem where every user needs to decode all messages not in its side information set. Although the optimal code-length may be the same in both centralized and decentralized settings, the actual optimal codes are not. For the decentralized PICOD, sparse Maximum Distance Separable (MDS) codes and vector linear index codes are used (as opposed to scalar linear codes). 
\end{abstract}

\section{Introduction}

\subsection{Motivation}
\label{sub:motivation}


Index coding (IC), first proposed when considering satellite communication~\cite{index_coding_original}, is a simple model to study the impact of message side information at the receivers in broadcast communication networks.
The IC consists of one transmitter with $m$ independent messages to be delivered to $n$ users through an error-free broadcast link. Each user has some messages as side information available to it and needs to reliably decode some messages that are not in its side information set; the desired messages for each user are pre-determined. In IC, one asks what is the minimum number of transmissions (i.e., minimum code-length) such that every user is able to decode its desired messages successfully. 
%
In this paper we are interested in the {\it decentralized pliable index coding problem}, which is motivated by two variants of IC: Pliable Index CODing (PICOD), and decentralized IC. 

The \emph{PICOD} problem is motivated by the flexibility in choosing the desired messages for the users in some practical scenarios, such as online advertisement systems.
Firstly proposed in~\cite{BrahmaFragouli-IT1115-7254174}, 
in the PICOD$(\cardi)$ there is a single transmitter, with $m$ message, and $n$ users, with message side information, which are connected via an error-free rate-limited broadcast channel, as in IC. Different from IC, in the PICOD$(\cardi)$ the desired messages at the users are not pre-determined and each user is satisfied whenever it can decode \emph{any} $\cardi$ messages not in its side information set. This provides the transmitter more encoding opportunities, as it now encodes based on its own choice of desired messages for the users. 
The goal in the PICOD$(\cardi)$ is to find the assignment of desired messages that leads to the smallest possible code-length.

The \emph{decentralized IC} is motivated by 
 peer-to-peer and ad-hoc network, where a central controller / transmitter does not exist and instead communication occurs among peers / users.
The decentralized IC can be seen as a special case of the distributed IC~\cite{capacity_thm_for_distributed_ic}.
In the distributed IC with $m$ messages, there are $2^m-1$ servers; each sender has knowledge of a unique subset of the message set (and can thus only encode based on its local knowledge) and is connected to the users through a separate error-free rate-limited link.
The decentralized IC is thus a distributed IC where there are as many servers as users, and each server has the same message knowledge as one of the users.
The goal for the decentralized IC is to determine the smallest number of channel uses such that all users are able to decoded their desired messages. 

The \emph{decentralized PICOD} proposed in this paper is a combination of the (centralized) PICOD and the decentralized IC, 
namely, a central transmitter with knowledge of all messages is not present, and instead users share among themselves massages that can only depend on their local side information set. 
The decentralized PICOD problem is motivated by  coded cooperative data exchange and distributed storage~\cite{d2d_caching}. 

\subsection{Past Work}
\label{sub:past_work}
Several achievable schemes have been proposed for PICOD, based on scalar linear codes;
the results of~\cite{BrahmaFragouli-IT1115-7254174,polytime_alg_picod} show an exponential code-length reduction for PICOD compared to IC.



For converse results, the optimal code-length under the restriction that the transmitter can only use linear schemes was shown in~\cite{BrahmaFragouli-IT1115-7254174} for the \emph{oblivious} PICOD$(\cardi)$, where the transmitter only knows the size of the side information at the users.
In~\cite{consecutive_picod}, we used techniques based on combinatorial design to prove tight converse bounds for some \emph{complete--$S$} PICOD$(\cardi)$ (see next for a formal definition) problems that generalize of the oblivious class; in those cases we showed the information theoretic optimality of scalar linear codes.


The multi-sender IC has been studied in~\cite{the_single_uniprior_index_coding_problem}, where the focus was on the ``single uniprior'' case (where users have only one single message as side information). The general multi-sender IC, or distributed IC, was investigated in~\cite{capacity_thm_for_distributed_ic}, where 
converse bounds (leveraging the submodularity of entropy) and achievable bounds (based on composite IC coding) were proposed; those bounds were numerically verified to match for the case of symmetric rate and symmetric link capacities in all settings with no more than four messages; the use of those bounds in general settings is however problematic because the number of variables involved is exponential in the number of servers (thus double exponential in the number of messages).

\subsection{Contributions}
\label{sub:contributions}
In this paper we derive tight information theoretic converse bounds (i.e., no restrictions on the class of codes used by the users) for two classes of decentralized PICOD$(\cardi)$ problems, 
namely:
(i)  the \emph{complement-consecutive complete}-$S$ PICOD$(\cardi)$, and
(ii) the \emph{consecutive complete}-$S$ PICOD$(\cardi)$.
The complete--$S$ PICOD$(\cardi)$, where $S$ is a subset of $[0:m-\cardi]$ (where $m$ is the number of messages at the transmitter and $t$ the number of messages to be decoded by each user), is a system where all side information sets / users with size indexed by $S$ are present. We say that $S$ is \emph{consecutive} if $S=[\smin: \smax]$ for some $0\leq \smin\leq \smax\leq m-\cardi$, and  \emph{complement-consecutive} if $S=[0:m-\cardi]\backslash[\smin:\smax]$ for some $0 < \smin\leq \smax < m-\cardi$. We characterized the optimal code-length in those cases in~\cite{consecutive_picod} for the centralized PICOD$(\cardi)$ case. Here, we examine the decentralized case.


Trivially, a centralized PICOD$(\cardi)$ has optimal block-length no larger than that of the corresponding decentralized problem (because a centralized transmitter can mimic any decentralized transmission scheme). 
In this work, we thus start by analyzing the decentralized version of those PICOD$(\cardi)$ problems whose optimal code-length we characterized in~\cite{consecutive_picod}, and use our tight past result as a ``trivial centralized converse bound.''
\emph{Surprisingly, we show that such a ``trivial centralized converse bound'' is tight whenever the decentralized PICOD$(\cardi)$ remains indeed pliable.} More precisely, we show that by using vector linear codes (in contrast to the simple linear scalar schemes that are optimal in the corresponding centralized setting~\cite{consecutive_picod}) we can achieve the ``trivial centralized converse bound'' except for the case where the problem parameters are such that every user must decode all the messages that are not in its side information set, that is, the problem becomes a multicast IC.


\subsection{Paper Organization}
\label{sub:paper_organization}
The rest of the paper is organized as follows: 
Section~\ref{sec:system_model} introduces the system model and related definitions;
Section~\ref{sec:main_results} summarizes our main contributions;
Section~\ref{sec:consescutive_s} provides the proof for consecutive complete--$S$ PICOD$(\cardi)$ and
Section~\ref{sec:complement_consecutive_s} for complement-consecutive complete--$S$ PICOD$(\cardi)$.
Section~\ref{sec:conclusion} concludes the paper.

\subsection{Notation}
\label{sub:notation}
Throughout the paper we use capital letters to denote sets, calligraphic letters for family of sets, and lower case letters for elements in a set. 
For integers $1 \leq a_1\leq a_2 $ we let $[a_1:a_2] := \{a_1,a_1+1,\ldots,a_2\}$, and $[a_2]:=[1:a_2]$.
A capital letter as a subscript denotes set of elements whose indices are in the set, i.e., $W_A:=\{w_a : w_a\in W, a\in A\}$.
For two sets $A$ and $B$, $A\setminus B$ is the set that consists all the elements that are in $A$ but not in $B$.



\section{System Model}
\label{sec:system_model}

%

A decentralized PICOD$(\cardi)$ system consists of:
(i) $n\in\mathbb{N}$ users and no central transmitter. The user set is denoted as $U := \left\{ u_{1},u_{2},\ldots,u_{n} \right\}$.
(ii) $m\in\mathbb{N}$ independent and uniformly distributed binary messages of $\mbit \in \mathbb{N}$ bits each. The message set is denoted as $W := \left\{ w_{1},w_{2},\ldots,w_{m} \right\}$.
(iii) User $u_i$ 
knows the messages indexed by its side information set $A_i\subset [m]$, $i\in[n]$. The collection of all side information sets is denoted as $\mathcal{A} := \{A_{1},A_{2},\ldots,A_{n}\}$, which is assumed globally known at all users. Note that for a decentralized PICOD problem to have a solution, one must have $\cup_{i=1}^{n} A_i \supset  A_j, \forall j\in [n]$, that is, for every user there must be an unknown message that is in the side information set of some other users.
(iv) An error-free broadcast link is shared among all users and allows one user to transmit while all the remaining users receive.
(v) The codeword $x^{\mbit\ell} := (x^{\mbit\ell_1}, x^{\mbit\ell_2}, \dots, x^{\mbit\ell_n})$ is eventually received by all users,
where $\ell := \sum_{j\in[n]} \ell_j$ and
\begin{align*}
     x^{\mbit\ell_j} := \mathsf{ENC}_{j} (W_{A_j}, \mathcal{A}), \ \forall j\in [n],
\end{align*} 
is the encoding function at user $u_j$.
%
(vi) 
The decoding function for user $u_j$ is 
\begin{align*}
    \{\widehat{w}^{(j)}_{1},\dots,\widehat{w}^{(j)}_{\cardi}\} 
    := \mathsf{DEC}_j(W_{A_j},x^{\ell \mbit}), \ \forall j\in [n].
\end{align*}
(vii) A code is \emph{valid} if and only if every user can successfully decode at least $\cardi$ messages not in its side information set, i.e., the decoding functions $\{ \mathsf{DEC}_j, \forall j\in[n]\}$ are such that
\begin{align*}
    \Pr[ & \exists \{d_{j,1},\dots,d_{j,\cardi}\}\cap A_j=\emptyset :  \\
    & \{\widehat{w}^{(j)}_{1},\dots,\widehat{w}^{(j)}_{\cardi}\}\neq \{{w}_{d_{j,1}},\dots,{w}_{d_{j,\cardi}}\}]  \leq \epsilon,
\end{align*}
for some $\epsilon\in(0,1)$. For a valid code, $\{\widehat{w}^{(j)}_{1},\dots,\widehat{w}^{(j)}_{\cardi}\} = \{{w}_{d_{j,1}},\dots,{w}_{d_{j,\cardi}}\}$ is called the \emph{desired message set} for user $u_j, \ j\in[n]$. The indices of the desired messages are denoted as $D_j:=\{d_{j,1},\dots,d_{j,\cardi}\}$  where $D_j\cap A_j=\emptyset, \forall j\in [n]$.
The choice of desired messages for the users is denoted as $\mathcal{D}=\{D_1,D_2,\ldots\,D_n\}$.
(viii)  The goal is to find a valid code with minimum length, that is, to determine
\begin{align*}
    \ell^{\star}:= \min\{\ell : \text{$\exists$ a valid $x^{\mbit\ell}$ for some $\mbit$}\}. 
\end{align*}

In the following we shall focus on the decentralized \emph{complete--$S$} PICOD$(\cardi)$, for a given set $S\subseteq[0:m-\cardi]$.
In this complete--$S$ system, there are $n := \sum_{s\in S}\binom{m}{s}$ users, where no two users have the same side information set, i.e., all possible users with distinct side information sets that are subsets of size $s$ of the message set, for all $s\in S$, are present in the system. 
In particular, we focus on 
the \emph{consecutive complete}-$S$ PICOD$(\cardi)$ and
the \emph{complement-consecutive complete}-$S$ PICOD$(\cardi)$, where
we say that $S$ is \emph{consecutive} if $S=[\smin: \smax]$ for some $0\leq \smin\leq \smax\leq m-\cardi$ (i.e., $S$ contains consecutive integers, from $\smin$ to $\smax$), and  \emph{complement-consecutive} if $S=[0:m-\cardi]\backslash[\smin:\smax]$ for some $0 < \smin\leq \smax < m-\cardi$ (note that the set $S$ includes elements $0$ and $m-\cardi$). 
We characterized the optimal centralized code-length in those two cases in~\cite{consecutive_picod} and we examine here the decentralized version.
Note that $S=\{0\}$ is not considered since it violates the condition $\cup_{i=1}^{n} A_i \neq A_j, \forall j\in[n]$.

\section{Main Results}
\label{sec:main_results}
The main contributions of this paper are as follows. 

\begin{thm}[consecutive]
\label{thm:consecutive_s}
	For the decentralized complete--$S$ PICOD$(\cardi)$ with $m$ messages and $S=[\smin:\smax]$ for some $0\leq \smin \leq \smax \leq m-\cardi$, the optimal code-length is
	\begin{align}
	\ell^{\star}=\begin{cases}
		\frac{\binom{m}{m-\cardi}}{\binom{m}{m-\cardi}-1}  \cardi, \quad \smax=\smin=m-\cardi,\\
		\min\{\smax+\cardi, m-\smin\}, \quad \text{otherwise.}
	\end{cases}
	\label{eq:thm:consecutive_s}
	\end{align}
\end{thm}

\begin{thm}[complement-consecutive]
\label{thm:complement_consecutive_s}
	For the decentralized complete--$S$ PICOD$(\cardi)$ with $m$ messages and $S=[0:m-\cardi]\backslash [\smin:\smax] = [0:\smin-1]\cup[\smax+1:m-\cardi]$ for some $0 < \smin \leq \smax < m-\cardi$, the optimal code-length is
	\begin{align}
	\ell^{\star}=\min\{m,|S|+2\cardi-2\}.
	\label{eq:thm:complement_consecutive_s}
	\end{align}
\end{thm}

Before we give the proof details in Sections~\ref{sec:consescutive_s} and~\ref{sec:complement_consecutive_s}, few remarks are in order:
\begin{enumerate}

\item
Surprisingly, Theorem~\ref{thm:complement_consecutive_s} says that, for the same parameters of $(m,\cardi,\smin, \smax)$, the centralized and the decentralized settings have the same optimal code-length; similarly for Theorem~\ref{thm:consecutive_s}, except for the case $\smax=\smin=m-\cardi$.

\item
Having the same optimal code-length does not necessarily imply that the same code is optimal in both cases. In~\cite{consecutive_picod}, we showed that for the centralized setting simple \emph{scalar linear} codes are optimal; in particular, the central transmitter either sends $\ell^{\star}$ distinct messages one by one, or $\ell^{\star}$ random linear combinations of all the messages. Clearly, the former strategy can be implemented in a decentralized setting, but not the latter. In this case we show that \emph{vector linear} codes are necessary; in particular, our achievable scheme uses \emph{sparse Maximum Distance Separable (MDS) codes}.

\item
When necessary to distinguish the optimal code lengths of the centralized and decentralized setting, we shall use the notation $\ell^{\star,\text{cen}}$ and $\ell^{\star,\text{dec}}$, respectively. Note that $\ell^{\star,\text{dec}}=\ell^{\star}$ where $\ell^{\star}$ was defined in Section~\ref{sec:system_model}.

Among all PICOD cases studied in this work, the only case where the decentralized optimal code-length $\ell^{\star,\text{dec}}$ is strictly larger than the corresponding centralized optimal code-length $\ell^{\star,\text{cen}}$ is when $\smin=\smax=m-\cardi$.
	This is the only case in centralized PICOD$(\cardi)$ where $\ell^{\star,\text{cen}} = \cardi$.
	Since $\ell^{\star,\text{dec}}>\cardi$ for all decentralized PICOD$(\cardi)$ (as what is sent by a user is not useful for that user), our results show that for the consecutive and complete consecutive complete--$S$ PICOD$(\cardi)$, $\ell^{\star,\text{dec}} \neq \ell^{\star,\text{cen}}$ if only if $\ell^{\star,\text{cen}} = \cardi$; interestingly, this is also the case where the PICOD problem looses its pliability, that is, it reduces to an IC problem where every user needs to decode all messages not in its side information set.

\item
	Theorems~\ref{thm:consecutive_s} and~\ref{thm:complement_consecutive_s} can be extended to all the centralized complete--$S$ PICOD$(\cardi)$ that we have been solved in~\cite{consecutive_picod}, which are not reported here because of space limitations. In those cases too we obtained $\ell^{\star,\text{dec}} = \ell^{\star,\text{cen}}$ whenever $\ell^{\star,\text{cen}} \neq \cardi$; and $\ell^{\star,\text{dec}} = \frac{n}{n-1}\cardi$ if $\ell^{\star,\text{cen}} = \cardi$. An intriguing question is whether this holds true for all complete--$S$ PICOD$(\cardi)$, even those not solved by the technique in~\cite{consecutive_picod}.  Answering this question is part of ongoing work.



\item
The similar proof technique can show that for the decentralized PICOD$(1)$ where the network topology hypergraph is a circular-arc~\cite{consecutive_picod}, the optimal code-length is: $\ell^\star=2$ if 1-factor does not exists; or $\ell^\star=\frac{p}{p-1}$, where $p$ is the maximum size of the 1-factor of the network topology hypergraph. 
This serves a tight bound for the decentralized PICOD beyond the complete--$S$ case.
Finding tight bounds for the general decentralized PICOD$(\cardi)$ is one direction for future work.


\end{enumerate}

\section{Proof for Theorem~\ref{thm:consecutive_s}} 
\label{sec:consescutive_s}


We split the proof into sub cases.
For $\ell^{\star,\text{cen}} = \min\{\smax+\cardi, m-\smin\} < \cardi$ for the centralized consecutive complete--$S$ PICOD$(\cardi)$, in which case $\ell^{\star,\text{cen}}=\ell^{\star,\text{dec}}=\ell^{\star}$, we study separately the cases $\smax +\cardi \leq m-\smin$ (Section~\ref{sub:smax+t<m-smin}) and $\cardi < m-\smin < \smax +\cardi$ (Section~\ref{sub:m-smin<smax+t}).
The case $\ell^{\star,\text{cen}} = \cardi$ is studied in Section~\ref{sub:consecutive_smin=smax=m-t}, in which case $\ell^{\star,\text{cen}}< \ell^{\star,\text{dec}}=\ell^{\star}=\frac{\binom{m}{m-\cardi}}{\binom{m}{m-\cardi}-1}  \cardi$ and is only possible for $\smin=\smax=m-\cardi$.


\subsection{Case $\smax +\cardi \leq m-\smin $} 
\label{sub:smax+t<m-smin}
We send $\smax+\cardi$ messages, one at a time. 
This can be done in a decentralized setting since each message is in the side information set of at least one user. Therefore, such a user can transmit the message to the rest of the users in one channel use. This achievable scheme is optimal since $\smax+\cardi$ is the optimal code-length for the corresponding centralized setting. 
We thus conclude $\ell^{\star}=\smax+\cardi$ for $\smax +\cardi \leq m-\smin$.

\subsection{Case $\cardi < m-\smin < \smax +\cardi$}
\label{sub:m-smin<smax+t}

We show that in this case a decentralized scheme with $m-\smin$ transmissions can satisfy all users; being $m-\smin$ the optimal code-length for the corresponding centralized setting, such a scheme is thus optimal. In the centralized case, the optimal code involves $m-\smin$ linearly independent linear combinations of all the messages, or alternatively an MDS code; this is not a possible decentralized scheme because at most $\smax+\cardi$ messages can be used to produce a valid code (assuming that a user has side information set of size $\smax$ and sends after having decoded $\cardi$ messages).

In the rest of the paper, when describing achievable schemes, instead of working with messages and codewords in bits (as done in the description of the channel model in Section~\ref{sec:system_model}), we represent each message of $\mbit$~bits as one symbols in the finite field $\mathbb{F}_{2^\mbit}$. With an abuse of notation, we also let $x^{\ell}$ denote the codeword of length $\ell$ symbols from the finite field $\mathbb{F}_{2^\mbit}$, and where each symbol corresponds to a transmission by a user.
A linear code for the decentralized system is thus $x^{\ell} = \mathbf{G} w^m$, where $\mathbf{G}$ is the code generator matrix of size $\ell \times m$ and $w^{m}$ is the vector of length $m$ containing all the messages.



For a valid optimal decentralized linear code, we look for a matrix $\mathbf{G} = [\mathbf{C} , \mathbf{0}]$, where $\mathbf{0}$ is zero matrix of size $\ell^{\star} \times (m-\smax-\cardi)$ with $\ell^{\star}=m-\smin$, and where $\mathbf{C}$ is a matrix of size $\ell^{\star} \times (\smax+\cardi)$ that satisfies two conditions:
\begin{enumerate}
 	\item\label{item:condition1} [C\ref{item:condition1}]
	each row has at most $\smax$ non-zero elements, and
 	\item\label{item:condition2} [C\ref{item:condition2}]
	any submatrix of $p$ columns, with $\cardi \leq p \leq \ell^{\star}$, has rank $p$ / is full rank.
\end{enumerate} 
The reason for these conditions is as follows.
Each row of $\mathbf{G}$ is the encoding vector used by a user; C\ref{item:condition1} is because a user knows at most $\smax$ messages (in its side information set).
C\ref{item:condition2} is for successful decoding at the users; once the contribution of the messages in the side information set has been subtracted off from the code, each user sees a subset of the remaining messages encoded by a full rank submatrix of $p$ columns; the range of $p$ is because each user must decode at least $\cardi$ messages, thus $\cardi \leq p$, and at most all messages in the code that are not in the side information, thus $p \leq \ell^{\star}$.

Note that condition C\ref{item:condition2} is equivalent to require that all $\ell^\star \times \ell^\star$ submatrices of $\mathbf{G}$ are full rank. This is because any submatrix obtained by taking a subset of columns of a full rank square matrix is full rank. Therefore, instead of having to consider all possible sets of $p$ columns in condition C\ref{item:condition2}, we only look at submatrix of size $\ell^\star \times \ell^\star$, which is the so-called MDS-property of a linear code of dimension $\ell^\star$.
We show that the desired matrix $\mathbf{G}$ exists as a spare MDS code generator matrix for sufficiently large $\mbit$, that is, for a sufficiently large field size.

We now introduce the ``zero pattern'' matrix for the spare MDS code generator matrix.
The ``zero pattern'' matrix $\mathbf{Z}\in \{0,1\}^{(m-\smin) \times (\smax+\cardi)}$ of $\mathbf{C}$ is a matrix whose entry is $0$ if the corresponding entry in $\mathbf{C}$ is $0$, and $1$ otherwise. 
Consider the following $\mathbf{Z}=[z_{ij}]$
\begin{align*}
	z_{ij} = 
	\begin{cases}
		1, & \text{for $0 \leq i+j \Mod{(\smax+\cardi)} \leq \smax-1$,} \\
		0, & \text{otherwise.}
	\end{cases}
\end{align*}
Let $Z_i:=\{j\in [\smax+\cardi] : z_{ij} =0\}$ be the set of the zero entries in the $i$th row, $|Z_i|=\cardi, \forall i\in [m-\smin]$. 
Since $\smax+\cardi > m-\smin$, we have $Z_i\neq Z_j$, $i\neq j$. 
Therefore, all $Z_i$ are different ``shifted'' version of $Z_{1}$. In $\cap_{i\in P}Z_i$ there are $|P|-1$ ``shifts'', which reduce the size of the intersection by at least $|P|-1$.
We then have the inequality
\begin{align*}
	|P|+|\cap_{i\in P}Z_i |\leq |P|+\cardi-(|P|-1)=t+1\leq \ell^{\star},
\end{align*}
which is known as the ``MDS condition''(which is sufficient for the existence of an MDS generator matrix over some finite field~\cite{S18}). Therefore, there exists a matrix $\mathbf{C}$ that satisfies conditions C\ref{item:condition1} and C\ref{item:condition2} with the specified ``zero pattern'' $\mathbf{Z}$. From~\cite{S18}, a finite field of size $m-\smin+\smax+\cardi-1$ suffices.
Since $\mathbf{G}$ satisfies condition C\ref{item:condition1}, this code thus can be generated in a distributed way.

After receiving the codewords of length $\ell^{\star}=m-\smin$, user $u_i$ subtracts off the messages in its side information set $A_i$ and is left with a linear code for the messages $W_{[\smax+\cardi]\setminus A_i}$.
Condition C\ref{item:condition2} guarantees that all users can decode at least $\cardi$ messages that are not in their side information. 
Therefore all users can be satisfied by this code of length $m-\smin$. 
This concludes the proof for this case.

\subsection{Case $\smin=\smax=m-\cardi$}
\label{sub:consecutive_smin=smax=m-t}

Let $s:=\smin=\smax=m-\cardi$.
This is the case where the ``trivial centralized converse bound'' $\ell^{\star,\text{cen}}=\min\{m-s, s+\cardi\}=\cardi \leq \ell^{\star}$ is not tight, and for which we want to show 
$\ell^{\star}=\frac{\cardi\binom{m}{s}}{\binom{m}{s}-1}> \cardi=\ell^{\star,\text{cen}}$.
%
In this case, the decentralized PICOD$(\cardi)$ becomes an actual multicast decentralized IC problem, we must show both achievability and converse.

\subsubsection{Converse}
\label{ssub:consecutive_converse_new}
An intuitive explanation for the converse proof is as follows. 
The $n := {m \choose s}$ users in the system are symmetric, i.e., by relabeling the messages we can swap any pair of users. Therefore all users have the same ``chance'' $1/n$ to be the one who sends part of the overall codeword $x^{\ell}$. In the decentralized setting, the part of $x^{\ell}$ sent by a user is generated based on its own side information set, and such a transmission cannot benefit the transmitting user.
Therefore, at most a fraction $\frac{n-1}{n}$ of $x^{\ell}$ can be useful for each user. 
Since each transmission can convey at most one message, in order to let each user decode at least $\cardi$ messages, the total number of transmissions satisfies $\frac{n-1}{n} \ell \geq t$.

We next provide the formal proof for the converse.
Let $\ell_i \mbit$ be the number of bits sent by user $u_i, i\in[n]$, and $x^{\mbit\ell} := (x^{\mbit\ell_1}, x^{\mbit\ell_2}, \dots, x^{\mbit\ell_n})$ be the overall codeword used for decoding by the users, with $\ell := \sum_{i\in [n]}\ell_i$.
With an abuse of notation, let $x^{(\ell-\ell_i) \mbit}$ indicate the bits in the transmitted codeword $x^{\ell \mbit}$ that were not sent by user $u_i, i\in[n]$.
 
By Fano's inequality, 
with $\lim_{\mbit\to\infty} \epsilon_\mbit = 0$, we have 
\begin{align*}
	\ell \mbit \epsilon_\mbit  
	&\geq H(W_{D_i}| x^{\ell \mbit}, W_{A_i})  
	= H(W_{D_i}| x^{(\ell-\ell_i) \mbit}, W_{A_i})  \\
	&= H(W_{D_i}|W_{A_i})- I(W_{D_i} ; x^{(\ell-\ell_i) \mbit}|W_{A_i})  \\
	&= H(W_{D_i}) - I(W_{D_i} ; x^{(\ell-\ell_i) \mbit}|W_{A_i}),
\end{align*}
Therefore, for $\forall i\in [n]$, we have
\begin{align*}
	(\ell-\ell_i) \mbit 
	&\geq H(x^{(\ell-\ell_i) \mbit})  
	\geq H(x^{(\ell-\ell_i) \mbit}|W_{D_i})  \\
	&\geq I(W_{D_i} ; x^{(\ell-\ell_i) \mbit}|W_{A_i})  \\
	&\geq H(W_{D_i}) - \ell \mbit\epsilon_\mbit  
	\geq \cardi \mbit - \ell \mbit\epsilon_\mbit  
\end{align*}
and therefore, for large enough $\mbit$, by summing the above inequalities we obtain the converse bound
\begin{align}
	\label{eq:converse_bound_smax=smin=m-t}
	\ell \geq \frac{n\cardi}{n-1} = \frac{\binom{m}{s}}{\binom{m}{s}-1}  \cardi.
\end{align}

\subsubsection{Achievability}
\label{ssub:consecutive_achievability_new}
The achievability involves message splitting and random linear coding.
%
i.e., we use a vector linear code, in contrast to the scalar linear code used in Section~\ref{sub:m-smin<smax+t}.

We split each message into $\fspli$ sub-messages, $w_i=[w_{i,1}, w_{i,2}, \dots, w_{i,\fspli}], i\in[m]$. 
The size of the sub-message is $\mbit/\fspli$ bits, which is assumed to be an integer. The parameter $\fspli$
will be appropriately chosen later.
Each sub-message is thus on the finite field $\mathbb{F}_{2^{\mbit/\fspli}}$.
Each user uses $\ell^{\prime}=\frac{\fspli\ell}{n}$ sub-timeslots (as the messages are split into $\fspli$ pieces, the time slots are split into $\fspli$ pieces as well) to transmit.  
In each sub-timeslot the user transmits a linear combination of all the sub-messages it has in its side information set, i.e., at sub-timeslot $h$, user $u_i$ transmits $\sum_{g\in A_i, j\in[\fspli]} a_{gj}(h) w_{g,j}$, where the coefficients $a_{gj}(h)$ are on $\mathbb{F}_{2^{\mbit/\fspli}}$.
The linear code has generator matrix $\mathbf{G}$ consisting of $a_{gj}(h)$ for $g\in[m], j\in[\fspli], h\in [\fspli\ell]$, and of size $n \ell^{\prime} \times m\fspli$. Each row of $\mathbf{G}$ has at most $s\fspli$ nonzero entries. 

For each user, among all $n \ell^{\prime}$ sub-timeslots, only $(n-1)\ell^{\prime}$ are useful for its decoding since the other $\ell^{\prime}$ sub-timeslots are transmitted by itself. 
Therefore, we choose $\ell^{\prime}$ and $\fspli$ such that 
$$
(n-1)\ell^{\prime}=(m-s)\fspli, \ n = {m \choose  s}, \ \ell^{\prime}=\frac{\fspli\ell}{n}.
$$ 
For each user, the submatrix of $\mathbf{G}$ corresponding to what all other users have sent needs to be a full rank square matrix of size $(n-1)\ell^{\prime} \times (m-s)\fspli$ so that each user can successfully decode.
In other words, every submatrix of $\mathbf{G}$ formed by $(m-s)\fspli$ columns is full rank.
Similarly to the proof in Section~\ref{sub:m-smin<smax+t}, the ``MDS condition'' on its zero-pattern matrix is as follows
\begin{align*}
	|P|+|\cap_{i\in P} Z_i| 
	& \leq |P|+\left((m-s)-(\ceil{\frac{|P|}{\ell^{\prime}}}-1)\right)\fspli \\
	& \leq n \ell^{\prime} +|P|-\ell^{\prime} -\frac{|P|-\ell^{\prime}}{\ell^{\prime}}\fspli 
	\leq n \ell^{\prime}.
\end{align*}
Therefore the proposed code generator matrix $\mathbf{G}$ exists for some large enough $\mbit$.
By this scheme
each user decodes all the $(m-s)\fspli$ sub-messages that are not in its side information. 

The total number of transmissions by this scheme is
\begin{align}
	\label{eq:ach_bound_smax=smin=m-t}
	\ell 
	= \frac{\ell^{\prime}}{\fspli} n
	= \frac{1}{\fspli}\frac{\fspli (m-s)}{n-1} n 
	= \frac{n\cardi}{n-1},
\end{align}
which coincides with the converse bound in~\refeq{eq:converse_bound_smax=smin=m-t}.
Therefore the achievability scheme is information theoretically optimal.



\section{Proof for Theorem~\ref{thm:complement_consecutive_s}} 
\label{sec:complement_consecutive_s}
Also for this decentralized complement-consecutive complete--$S$ PICOD$(\cardi)$,
where $S=[0:m-1]\setminus [\smin:\smax]$ for some $0<\smin \leq \smax<m-\cardi$,
we need to show a decentralized achievable scheme that meet the ``trivial centralized converse bound.'' 

In the centralized case, the achievable scheme consists of two scalar linear codes:
one to serve all the users with side information of size in $[0:\smin-1]$, and 
the other to serve all the users with side information of size in $[\smax+1:m-\cardi]$. 
Also for the decentralized scheme, we separate the users into these two groups: 
$U_1 = \{u_i : |A_i|\in [0:\smin-1]\}$ and $U_2 = \{u_i : |A_i|\in [\smax+1: m-\cardi]\}$.
The analysis of the achievability scheme is divided into two parts: $\smin-1+\cardi<\smax+1=m-\cardi$ and the rest.

\subsection{Case $\smin-1+\cardi<\smax+1=m-\cardi$} 
\label{sub:case_1_complement_consecutive}
	In this case the decentralized scheme is different from the centralized one. 
	This is because $U_2$ in this case represents a consecutive complete--$S$ case discussed in Section~\ref{sub:consecutive_smin=smax=m-t}, where the centralized converse bound is not tight. 
	Therefore, we can not treat the problem of serving the users in $U_1$ and $U_2$ as two independent subproblems, as the scheme does in centralized case. 
	The achievability scheme takes two steps:
	\begin{itemize}
		\item{Step~1:} Send messages $W_{[\smin-1+\cardi]}$ one by one.
		All users in $U_1$ are satisfied.
		$\smin-1+\cardi\geq \cardi$ messages are sent in this step. Since  all users in $U_2$ have side information sets of size $\smax+1=m-\cardi$, there exists at least one user in $U_2$ that has been satisfied in the first step.
		\item{Step~2:} 
		The user in $U_2$ that was satisfied in Step~1 has the knowledge of all messages
		and can thus act as the centralized transmitter of the centralized PICOD$(\cardi)$~\cite{consecutive_picod}, sending $\cardi$ linearly independent linear combinations of all messages.
		Since all users in $U_2$ have $\cardi$ messages not in the side information, by having $\cardi$ linear independent linear combinations of all messages, all users in $U_2$ are satisfied.
	\end{itemize}
	It thus takes $\smin-1+\cardi+\cardi=|S|+2\cardi-2$ number of transmissions to satisfy all users.

\subsection{Other Case} 
\label{sub:other_case_complement_consecutive}
	The achievable scheme in Section~\ref{sub:smax+t<m-smin} satisfies the users in $U_1$ by using $\smin-1+\cardi$ transmissions. The achievable scheme in Section~\ref{sub:m-smin<smax+t} satisfies the users in $U_2$ by using $m-(\smax+1)$ transmissions. Therefore, the total number of transmissions is $\smin-1+\cardi+m-\smax-1=|S|+2\cardi-2$.

Note that $\ell=m$ is a trivially achievable number of transmissions for the decentralized setting as well, we conclude for decentralized complement-consecutive complete--$S$ PICOD$(\cardi)$ the optimal number of transmissions is $\ell^{\star} = \min\{m, |S|+2\cardi-2\}$, which is the same as the centralized setting.

\section{Conclusion} 
\label{sec:conclusion}
In this paper we introduced and found the capacity of some  decentralized complete--$S$ PICOD$(\cardi)$ problems.
For most cases we found that the optimal code-length for the decentralized setting is the same as for the centralized one.
Among the cases we have explored, we found that when all users request all the messages that are not in their side information set then the decentralized PICOD has a strictly larger optimal code-length then the centralized one. 
%
Whether there are other cases where the centralized and decentralized settings have the same code-length, and if there is a fundamental connection between lack of ``pliability'' and different code-lengths between centralized and decentralized settings, is part of ongoing work.


\bibliographystyle{IEEEtranS}
\bibliography{refs}

\end{document}